\RequirePackage{lineno} 

\documentclass[12pt,onecolumn,prc,showpacs,preprintnumber,amsmath,
floatfix,amssymb]{revtex4}
\usepackage{epsfig}
\usepackage{graphicx}
\usepackage{dcolumn}
\usepackage{bm}
\def\var{\mbox{\boldmath $\varepsilon$}}
\def\r{\mbox{{\bf  r}}}
\def\p{\mbox{\boldmath $p$}}
\def\q{\mbox{\boldmath $q$}}
\def\k{\mbox{\boldmath $k$}}
\def\t{\mbox{\boldmath $t$}}
\allowdisplaybreaks[4]

\begin{document}
\setpagewiselinenumbers

\title{Reduced cross sections of electron and neutrino charged current
  quasielastic scattering on nuclei }
\author{A.~V.~Butkevich}
\affiliation{Institute for Nuclear Research,
Russian Academy of Sciences, Moscow 117312, Russia\\}
\date{\today}
\begin{abstract}

  The semi-exclusive averaged reduced cross sections for (anti)neutrino charged
  current quasi-elastic scattering on carbon, oxygen, and argon are analyzed
  within the relativistic distorted wave impulse approximation. We found that
  these cross sections as functions of missing nucleon energy are similar to
  those of electron scattering and are in agreement with electron scattering
  data for three nuclei. The difference between the electron and neutrino
  cross sections can be attributed to Coulomb distortion on the electron wave
  function. The averaged reduced cross sections depend slowly upon incoming
  lepton energy. The approach presented in this paper provide novel constraints
  on nuclear models of quasi-elastic neutrino-nucleus scattering and can be
  easily applied to test spectral functions and final state interactions,
  employed in neutrino event generators.

\end{abstract}

\maketitle

\section{Introduction}

For current~\cite{NOvA1, T2K} and future~\cite{DUNE, HK2T, SBN}
accelerator-based neutrino experiments the primary physics goals are measuring
the lepton CP violation phase, determing neutrino mass ordering and testing the
three flavor paradigm. In these experiments to evaluate the oscillation
parameters, the probabilities of neutrino oscillations as functions of neutrino
energy are measured. The neutrino beams are not monoenergetic and have broad
distributions that range from tens of MeVs to a few GeVs. The accuracy to which
neutrino oscillation parameters can be extracted depends on the ability of
experiments to determine the individual energy of detected neutrino.  

Measurements at neutrino energy 1 GeV are critical for the T2K~\cite{T2K} and HK
~\cite{HK2T} programs, which are carbon and water (oxygen) detectors as well as
for the SBN (argon)~\cite{SBN} program. 
Measurements from 1 to 2 GeV are important for the 
NOvA (carbon, chlorine)~\cite{NOvA1} experiment, and measurements spanning from
1 to 10 GeV critical for the DUNE (argon)~\cite{DUNE} program. At the GeV-
scale neutrino energies the neutrino can interact with a nucleus through a wide
range of reaction channels. These include the charged-current (CC) quasielastic
 (QE) scattering, two-body meson exchange current (MEC) channels, resonance
production and deep inelastic scattering.

The incident neutrino energy is reconstructed using kinematic or
calorimetric methods. At energy about 1 GeV, where the CCQE scattering is
dominant, the incoming neutrino energy can be derived from lepton kinematics
alone. The calorimetric method relies not only on the visible energy measured
in the detector, but also on the models of the neutrino-nucleus interactions
that are implemented in neutrino event generators. In addition the
neutrino-nucleus scattering model is critical for obtain background estimates,
and for correct extrapolations of the near detector constraints to the far
detector in analyses aimed at determing the neutrino oscillation parameters.

Unfortunately, due to wide range of neutrino energy beams and poor statistics
available from current experiments, it is very difficult to measure
differential neutrino-nucleus cross sections for specific energies and to test
beam energy reconstruction techniques. On the theoretical side, many studies
have been presented aiming at improving our knowledge on lepton-nucleus
interaction~\cite{BAV1, BAV2, Martini1, Martini2, Nieves1, Nieves2, BAV4,
  Martini3, Simo, Megias1, Megias2, Megias3, Rocco, BAV5, BAV6, Gon1, Gon2,
  Gon3, BAV7, BAV8, BAV9, Kim}. However, it is extremely challenging to provide
reliable and consistent predictions for the diversity of processes that can take
place in the energy range covered by the neutrino beams. Various contributions
to the cross sections can significantly overlap with each other making it
difficult to identify, diagnose and remedy shortcoming of nuclear models.

While electron and neutrino interactions are different at the primary vertex,
many underlying physics process in the nucleus are the same, and
electron scattering data collected with precisely controlled kinematics (initial
and final energies and scattering angles) and large statistic allows 
validation and improvement of the description of nuclear effects. There are a large body
of electron-scattering data on carbon and calcium and only a few data sets
available for scattering on argon.

All of the above reaction mechanisms are very similar for electrons and for
neutrinos. From the nuclear point of view the influence of nuclear medium
effects such as the nuclear ground state and interaction of the outgoing
nucleon with the residual nucleus can be expected to be largely the same for
electron as for neutrino-induced processes. We can exploit this similarity and
use electron scattering data with known beam energies to test the neutrino
energy reconstruction methods~\cite{CLAS} and interaction models. The vector part
of the electroweak interaction can be inferred directly from the electron
scattering data. Because electron and neutrino scattering are strongly linked
in theory, any model of neutrino interactions (vector+axial) should also be
able to reproduce electron (vector) interactions. A model unable to reproduce
electron measurements cannot be expected to provide accurate prediction for
neutrino cross sections.

It is therefore unsurprising that recent years have seen a plethora of analyses
of electron-scattering data to test the vector current part of the
lepton-nucleus interaction against existing inclusive electron scattering cross
sections for different target nuclei at several incident beam energies and
scattering electron angles. The relativistic distorted wave impulse
approximation (RDWIA), initially designed for description of exclusive
$(e,e'p)$
data~\cite{Pick, Udias, JKelly} and then adopted for neutrino reactions was
successfully tested against of inclusive $(e,e')$ data~\cite{BAV7, Gon3}. The
SuSAv2 model exploits the similarities between both interaction types to guide
the description of weak scattering process~\cite{Megias2, Megias3}. The utility
of validating neutrino events generators against inclusive electron scattering
data that they had not been tuned to was demonstrated in Refs.~\cite{Ankow,
  e4v1, NEUT, Dytman}.

Such inclusive reactions involve total hadronic cross sections and typically
are relatively insensitive to the details of the final nuclear states.
Rather simple models may yield cross sections that are not very different from
those found in the most sophisticated models. Typically, the inclusive
predictions using different models are rather similar and agree to about
10-20\%, but they cannot make predictions on both leptons and hadrons in final
states. The semi-exclusive $(l,l'p)$ lepton scattering process involves not the
total cross sections, but the specific asymptotic states and allows to test
more in detail the nuclear model. Microscopic and unfactorized models like
the RDWIA can be used to model both lepton-boson and boson-nucleus
vertexes in the same detail and compare the results to semi-exclusive observables.
The comparison of the results of the RDWIA approach and cascade models employed
in the neutrino event generators provides constraints on cascade models from
proton-nucleus scattering~\cite{Udias}.

The reduced cross section, obtained from the measured differential
semi-exclusive electron scattering cross section dividing on the kinematic
factor and the off-shell electron-proton cross section, can be identified with
the distorted spectral function. Final state interactions between the ejected
nucleon and the residual nucleus make the reduced cross sections depend upon
the initial and ejectile nucleon's momenta and angle between them
(depends upon momentum transfer). Thus, irrespective of the type of interaction
(electromagnetic or weak) the distorted spectral function is determined mainly
by the intrinsic properties of the target and the ejected nucleon interaction
with residual nucleus.

The purpose of the present work is calculation of the CCQE neutrino scattering
reduced cross sections averaged over phase space as functions of the missing
nucleon momentum and incoming neutrino energy, and comparison of them with ones
obtained from measurements of $(e,e'p)$ scattering
on carbon, oxygen and argon targets. The direct comparison of the spectral
functions used in the factorized approach in neutrino event generators to the 
measured reduced cross sections of the electron-nucleus scattering can provide
an additional test of the nuclear models employed in these generators.

The outline of this paper is the following. In Sec.II we introduce
the formalism needed to describe  the semi-exclusive lepton-nucleus
CCQE scattering process. The RDWIA model is briefly introduced in Sec.III.
Results of the calculations are presented in Sec.IV. Our conclusions are
summarized in Sec.V.

\section{Formalism of quasi-elastic scattering}

We consider the formalism used to describe electron and neutrino quasi-elastic
exclusive
\begin{equation}\label{qe:excl}
l(k_i) + A(p_A)  \rightarrow l^{\prime}(k_f) + N(p_x) + B(p_B),      
\end{equation}
scattering off nuclei in the one-photon (W-boson) exchange approximation. 
Here $l$ labels the incident lepton [electron or muon (anti)neutrino], and
$l^{\prime}$ represents the scattered lepton (electron or muon),
$k_i=(\varepsilon_i,\k_i)$ 
and $k_f=(\varepsilon_f,\k_f)$ are the initial and final lepton 
momenta, $p_A=(\varepsilon_A,\p_A)$, and $p_B=(\varepsilon_B,\p_B)$ are 
the initial and final target momenta, $p_x=(\varepsilon_x,\p_x)$ is the 
ejectile nucleon momentum, $q=(\omega,\q)$ is the momentum transfer carried by 
the virtual photon (W-boson), and $Q^2=-q^2=\q^2-\omega^2$ is the photon 
(W-boson) virtuality.

\subsection{CCQE lepton-nucleus cross sections}

In the laboratory frame the differential cross section for exclusive
electron ($\sigma ^{el}$) and (anti)neutrino ($\sigma ^{cc}$) CC scattering can
be written as
\begin{subequations}
\begin{align}
\frac{d^6\sigma^{el}}{d\varepsilon_f d\Omega_f d\varepsilon_x d\Omega_x} &=
\frac{\vert\p_x\vert\varepsilon_x}{(2\pi)^3}\frac{\varepsilon_f}{\varepsilon_i}
 \frac{\alpha^2}{Q^4} L_{\mu \nu}^{(el)}\mathcal{W}^{\mu \nu (el)}
\\                                                                  
\frac{d^6\sigma^{cc}}{d\varepsilon_f d\Omega_f d\varepsilon_x d\Omega_x} &=
\frac{\vert\p_x\vert\varepsilon_x}{(2\pi)^5}\frac{\vert\k_f\vert}
{\varepsilon_i} \frac{G^2\cos^2\theta_c}{2} L_{\mu \nu}^{(cc)}
\mathcal{W}^{\mu \nu (cc)},
\end{align}
\end{subequations}
 where $\Omega_f$ is the solid angle for the lepton momentum, $\Omega_x$ is the
 solid angle for the ejectile nucleon momentum, $\alpha\simeq 1/137$ is the
fine-structure constant, $G \simeq$ 1.16639 $\times 10^{-11}$ MeV$^{-2}$ is
the Fermi constant, $\theta_C$ is the Cabbibo angle
($\cos \theta_C \approx$ 0.9749), $L^{\mu \nu}$ is the lepton tensor, and
 $\mathcal{W}^{(el)}_{\mu \nu}$ and $\mathcal{W}^{(cc)}_{\mu \nu}$ are
correspondingly the electromagnetic and weak CC nuclear tensors.

For exclusive reactions in which only a single discrete state or a narrow
resonance of the target is excited, it is possible to integrate over the
peak in missing energy and obtain a fivefold differential cross section of
the form
\begin{subequations}
\begin{align}
\label{cs5:el}
\frac{d^5\sigma^{el}}{d\varepsilon_f d\Omega_f d\Omega_x} &= R
\frac{\vert\p_x\vert\tilde{\varepsilon}_x}{(2\pi)^3}\frac{\varepsilon_f}
{\varepsilon_i} \frac{\alpha^2}{Q^4} L_{\mu \nu}^{(el)}W^{\mu \nu (el)}
\\                                                                       
\label{cs5:cc}
\frac{d^5\sigma^{cc}}{d\varepsilon_f d\Omega_f d\Omega_x} &= R
\frac{\vert\p_x\vert\tilde{\varepsilon}_x}{(2\pi)^5}\frac{\vert\k_f\vert}
{\varepsilon_i} \frac{G^2\cos^2\theta_c}{2} L_{\mu \nu}^{(cc)}W^{\mu \nu (cc)},
\end{align}
\end{subequations}
where $R$ is a recoil factor
  \begin{equation}\label{Rec}
R =\int d\varepsilon_x \delta(\varepsilon_x + \varepsilon_B - \omega -m_A)=
{\bigg\vert 1- \frac{\tilde{\varepsilon}_x}{\varepsilon_B}
\frac{\p_x\cdot \p_B}{\p_x\cdot \p_x}\bigg\vert}^{-1},                    
\end{equation}
$\tilde{\varepsilon}_x$ is solution to equation
$
\varepsilon_x+\varepsilon_B-m_A-\omega=0,
$
where $\varepsilon_B=\sqrt{m^2_B+\p^2_B}$, $~\p_B=\q-\p_x$ and $m_A$ and $m_B$
are masses of the target and recoil nucleus, respectively. Note, that missing
momentum is $\p_m=\p_x-\q$ and missing energy $\var_m$ is defined by
$\var_m=m + m_B -m_A$.

All information about the nuclear structure and effects of final-state
interaction (FSI) between the ejectile nucleon and residual nucleus is 
contained in the electromagnetic and weak CC hadronic tensors,
$W^{(el)}_{\mu \nu}$ and $W^{(cc)}_{\mu \nu}$, which are given by the bilinear 
products of the transition matrix elements of the nuclear electromagnetic or CC 
operator $J^{(el)(cc)}_{\mu}$ between the initial nucleus state 
$\vert A \rangle $ and the final state $\vert B_f \rangle$ as
\begin{eqnarray}
W^{(el)(cc)}_{\mu \nu } &=& \sum_f \langle B_f,p_x\vert                     
J^{(el)(cc)}_{\mu}\vert A\rangle \langle A\vert
J^{(el)(cc)\dagger}_{\nu}\vert B_f,p_x\rangle,              
\label{W}
\end{eqnarray}
where the sum is taken over undetected states.

In the exclusive reaction (\ref{qe:excl}) the outgoing lepton and proton are
detected and the exclusive lepton scattering cross sections (\ref{cs5:el}) and
 (\ref{cs5:cc}) in terms of response functions can be written as
\begin{widetext}
\begin{subequations}\label{cs5:R}
\begin{align}
\frac{d^5\sigma^{el}}{d\varepsilon_f d\Omega_f d\Omega_x} &=
\frac{\vert\p_x\vert\tilde{\varepsilon}_x}{(2\pi)^3}\sigma_M
R\big(V_LR^{(el)}_L + V_TR^{(el)}_T
 +V_{LT}R^{(el)}_{LT}\cos\phi + V_{TT}R^{(el)}_{TT}\cos 2\phi\big),
\\                                                                       
\frac{d^5\sigma^{cc}}{d\varepsilon_f d\Omega_f d\Omega_x} &=
\frac{\vert\p_x\vert\tilde{\varepsilon}_x}{(2\pi)^5}G^2\cos^2\theta_c
\varepsilon_f \vert \k_f \vert R \big \{ v_0R_0 + v_TR_T
 + v_{TT}R_{TT}\cos 2\phi + v_{zz}R_{zz}
\notag \\
& +(v_{xz}R_{xz} - v_{0x}R_{0x})\cos\phi  
-v_{0z}R_{0z} + h\big[v_{yz}(R^{\prime}_{yz}\sin\phi + R_{yz}\cos\phi)        
\notag \\
& - v_{0y}(R^{\prime}_{0y}\sin\phi + R_{0y}\cos\phi) - v_{xy}R_{xy}\big]\big\},
\end{align}
\end{subequations}
\end{widetext}
where
\begin{equation}
\sigma_M = \frac{\alpha^2\cos^2 \theta/2}{4\varepsilon^2_i\sin^4 \theta/2} 
\end{equation}
is the Mott cross section and $h$ is +1 for positive lepton helicity and -1 for
negative lepton helicity. The coupling coefficient $V_k$ and $v_k$, the
expression of which are given in Ref.~\cite{BAV1} are kinematic
factors depending on the lepton's kinematics. The response functions $R_i$
are given in terms of components of the exclusive hadronic tensors~
\cite{BAV1} and depend on the variables
$(Q^2, \omega)$ or $(|\q|,\omega)$.

It is also useful define a reduced cross section 
\begin{equation}
\sigma_{red} = \frac{d^5\sigma^{(el)(cc)}}{d\varepsilon_f d\Omega_f d\Omega_x}
/K^{(el)(cc)}\sigma_{lN},                                                 
\end{equation}
where
$K^{el} = R {p_x\varepsilon_x}/{(2\pi)^3}$ and
$K^{cc}=R {p_x\varepsilon_x}/{(2\pi)^5}$
are phase-space factors for electron and neutrino scattering  and 
$\sigma_{lN}$ is the corresponding
elementary cross section for the lepton scattering from the
moving free nucleon normalized to unit flux. The reduced cross section is an
interesting quantity that can be  regarded as the nucleon momentum distribution
modified by FSI, i.e. as the distorted spectral function. Final-state
interactions make the reduced cross sections $\sigma_{red}({\p}_m, {\p}_x)$
depend upon ejectile momentum ${\p}_x$, angle between the initial and final
nucleon momentum and upon incident lepton energy.
These cross sections for (anti)neutrino  scattering off
nuclei are similar to the electron scattering apart from small differences at
low beam energy due to effects of Coulomb distortion of the incoming electron
wave function as shown in Refs.~\cite{BAV1, BAV2, BAV4}

The factorization approximation to the knockout cross section stipulates that
\begin{equation}
\frac{d^5\sigma^{(el)(cc)}}{d\varepsilon_f d\Omega_f d\Omega_x}=
K^{(el)(cc)}\times \sigma_{lN} \times  \sigma_{red}(\var_m, {\p}_m, {\p}_x)    
\end{equation}
This factorization implies that the initial nuclear sate and FSI effects are 
decoupled from leptonic vertex with preserved the correlations between the final
 lepton and nucleon.

 The reduced cross section as a function of missing momentum $p_m$, averaged
 over phase volume in $(\omega, \Omega_f, \phi)$ coordinates, where
 $\Omega_x = (\cos\theta_{pq}, \phi)$, can be written as
 \begin{equation}
 \label{avsec}   
\langle \sigma^{red}(p_m)\rangle = \frac{1}{V}
\int d\phi \int d\var_f \int d\Omega_f \frac{p_m}{p_x\vert\q \vert}R_c  
\sigma^{red}(\var_f,\Omega_f,p_m,\phi),
\end{equation}
where
 $p_m=\vert\p_m\vert,~ p_x=\vert\p_x\vert,~ \p_m=\p_x-\q$, and
\begin{subequations}
\begin{align}
\cos\theta_{pq} &= \frac{\p^2_x + \q^2 - \p^2_m}{2p_x\vert\q\vert},    
\\
R_c &= 1 + \frac{\varepsilon_x}{2p^2_x\varepsilon_B}
(\p^2_x +\q^2 - \p^2_m),
\\
  V &= \int d\phi \int d\var_f\int  
d\Omega_f \frac{p_m}{p_x\vert\q \vert}R_c.
\end{align}
\end{subequations}
Precise electron reduced cross sections data can be used to validate the
neutrino reduced cross sections (spectral functions) that are implemented in
neutrino generators. 
 
 \subsection{ Nuclear current}

Obviously, the determination of the response tensor $W^{\mu\nu}$ requires the
knowledge of the nuclear current matrix elements
in Eq.(\ref{W}). We describe the lepton-nucleon scattering in the impulse
approximation, assuming that the incoming lepton interacts with only one
nucleon, which is subsequently emitted. The nuclear current is written as the
sum of single-nucleon currents. Then, the nuclear matrix element
in Eq.(\ref{W}) takes the form
\begin{eqnarray}
 \label{Eq.11} 
\langle p,B\vert J^{\mu}\vert A\rangle &=& \int d^3r~ \exp(i\t\cdot\r)
\overline{\Psi}^{(-)}(\p,\r)
\Gamma^{\mu}\Phi(\r),                                               
\end{eqnarray}
where $\Gamma^{\mu}$ is the vertex function, $\t=\varepsilon_B\q/W$ is the
recoil-corrected momentum transfer, $W=\sqrt{(m_A + \omega)^2 - \q^2}$ is the
invariant mass, $\Phi$ and $\Psi^{(-)}$ are relativistic bound-state and
outgoing wave functions.
For electron scattering, most calculations use the CC2 electromagnetic vertex
function for a free nucleon \cite{deFor}
\begin{equation}
 \label{Gamma} 
\Gamma^{\mu} = F^{(el)}_V(Q^2)\gamma^{\mu} + {i}\sigma^{\mu \nu}\frac{q_{\nu}}
{2m}F^{(el)}_M(Q^2),                                                 
\end{equation}
where $\sigma^{\mu \nu}=i[\gamma^{\mu}\gamma^{\nu}]/2$, $F^{(el)}_V$ and
$F^{(el)}_M$ are the Dirac and Pauli nucleon form factors. Because the bound
nucleons are off shell, the vertex $\Gamma^{\mu}$ in Eq.(\ref{Gamma}) should be
taken for an off-shell nucleon. We employ the de Forest prescription for
the off-shell
vertex~\cite{deFor}
\begin{equation}
\tilde{\Gamma}^{\mu} = F^{(el)}_V(Q^2)\gamma^{\mu} + {i}\sigma^{\mu \nu}
\frac{\tilde{q}_{\nu}}{2m}F^{(el)}_M(Q^2),                            
\end{equation}
where $\tilde{q}=(\varepsilon_x - \tilde{E},\q)$ and the nucleon energy
$\tilde{E}=\sqrt{m^2+(\p_x - \q)^2}$ is placed on shell. We use the
approximation of~\cite{MMD} on the nucleon form factors. The Coulomb gauge
is assumed for the single-nucleon current. Although the experimental
analysis usually employ the de Forest CC1 prescription for $\sigma_{lN}$,
consistency requires that calculation of $\sigma_{red}$ to employ the
$\sigma_{lN}$ that corresponds to the current operator used in the RDWIA
calculations.

The single-nucleon charged current  has $V{-}A$
structure $J^{\mu (cc)} = J^{\mu}_V + J^{\mu}_A$. For a free nucleon vertex
function $\Gamma^{\mu (cc)} = \Gamma^{\mu}_V + \Gamma^{\mu}_A$ we use CC2
vector current vertex function
\begin{equation}
\Gamma^{\mu}_V = F_V(Q^2)\gamma^{\mu} + {i}\sigma^{\mu \nu}\frac{q_{\nu}}
{2m}F_M(Q^2)                                                         
\end{equation}
and the axial current vertex function
\begin{equation}
\Gamma^{\mu}_A = F_A(Q^2)\gamma^{\mu}\gamma_5 + F_P(Q^2)q^{\mu}\gamma_5.  
\end{equation}
Weak vector form factors $F_V$ and $F_M$ are related to corresponding
electromagnetic ones for proton $F^{(el)}_{i,p}$ and neutron $F^{(el)}_{i,n}$
by the hypothesis of conserved vector current (CVC)
\begin{equation}
F_i = F^{(el)}_{i,p} - F^{(el)}_{i,n}.                                 
\end{equation}
The axial $F_A$ and psevdoscalar $F_P$ form factors in the dipole
approximation are parameterized as
\begin{equation}
F_A(Q^2)=\frac{F_A(0)}{(1+Q^2/M_A^2)^2},\quad                          
F_P(Q^2)=\frac{2m F_A(Q^2)}{m_{\pi}^2+Q^2},
\end{equation}
where $F_A(0)=1.2724$, $m_{\pi}$ is the pion mass, and $M_A$ is the axial mass.
We use de Forest prescription for off-shell extrapolation of
$\Gamma^{\mu (cc)}$. Similar to electromagnetic current, the Coulomb gauge is
applied for the vector current $J_V$.

\section{Model}

The semi-exclusive differential and reduced cross sections for neutrino
scattering were studied in Refs.~\cite{BAV1, BAV2, BAV4, BAV8, BAV9},
using the relativistic shell model approach and taking into account the FSI
effects. A formalism for the $A(e,e^{\prime}N)B$ reaction that describes the
channel coupling in the FSI of $N+B$ system was developed in Ref.~\cite{JKelly}.

In this work the independent particle shell model (IPSM) is assumed 
for the nuclear structure. The model space for
$^{12}$C$(l,l^{\prime}N)$ consists of $1s_{1/2}$ and $1p_{3/2}$
nucleon-hole states in the $^{11}$B and $^{11}$C nuclei.
The model space for $^{16}$O$(l,l^{\prime}N)$ consists of $1s_{1/2}$, $1p_{3/2}$,
and $1p_{1/2}$ nucleon-hole states in the $^{15}$N and $^{15}$O nuclei.
The model space for $^{40}$Ar$(l,l^{\prime}N)$ consists of $1s_{1/2}$, $1p_{3/2}$,
$1p_{1/2}$, $1d_{5/2}$, $2s_{1/2}$, and $1d_{3/2}$ nucleon-hole states in the
$^{39}$Cl, and $1s_{1/2}$, $1p_{3/2}$, $1p_{1/2}$, $1d_{5/2}$, $2s_{1/2}$,
$1d_{3/2}$, and $1f_{7/2}$  nucleon-hole states in the $^{39}$Ar nuclei.
All states in these nuclei are regarded as discrete states even though their
spreading widths are actually appreciable.

In the independent particle shell model the relativistic bound-state function 
$\Phi$ in Eq.(\ref{Eq.11}) is obtained as the self-consistent solutions of a
Dirac equation, derived within a relativistic mean-field approach, from a
Lagrangian containing $\sigma$, $\omega$, and $\rho$ mesons~\cite{Serot}. The
nucleon bound-state functions were calculated by the TIMORA code~\cite{Horow} 
with the normalization factors $S(\alpha)$ relative to full occupancy of the
IPSM orbitals. According the RDWIA analysis of the JLab ${}^{12}$C$(e,e'p)$ data
~\cite{Dutta, Kelly1} $S(1p_{3/2})=84\%$, $S(1s_{1/2})=100\%$ and average factor
about $\approx 89\%$. We use also the following values of normalization factors
of ${}^{16}$O: $S(1p_{3/2})=66\%$, $S(1p_{1/2})=70\%$, and $S(1s_{1/2})=100\%$,
that were obtained in the RDWIA analysis of the JLab data~\cite{Fissum}.
From the RDWIA analysis~\cite{BAV4} of NIKHEF
data~\cite{Kramer1, Kramer2, Kramer3} follows that the occupancy of the
\begin{figure*}
  \begin{center}
    \includegraphics[height=9cm,width=12cm]{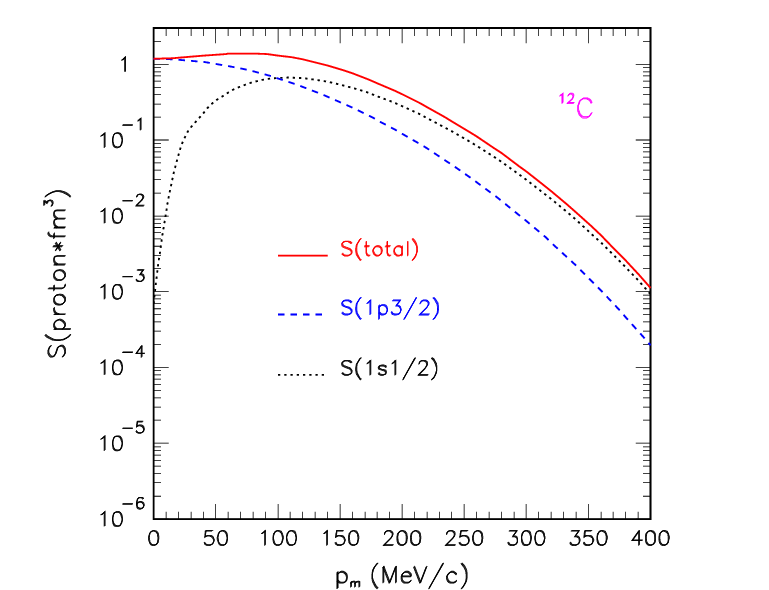}
  \end{center}
  \caption{\label{Fig1} Proton momentum distributions for the different single
    particle states in ${}^{12}$C nucleus. Also shown is the total proton
    momentum distribution (solid line).
}
\end{figure*}
\begin{figure*}
  \begin{center}
    \includegraphics[height=9cm,width=12cm]{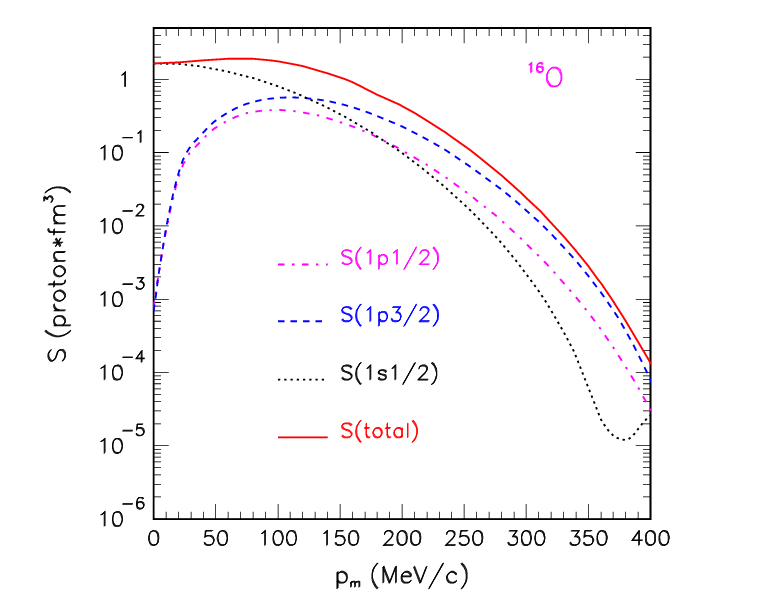}
  \end{center}
  \caption{\label{Fig2} Same as Fig.1 but in ${}^{16}$O nucleus.
}
\end{figure*}
\begin{figure*}
  \begin{center}
    \includegraphics[height=9cm,width=11cm]{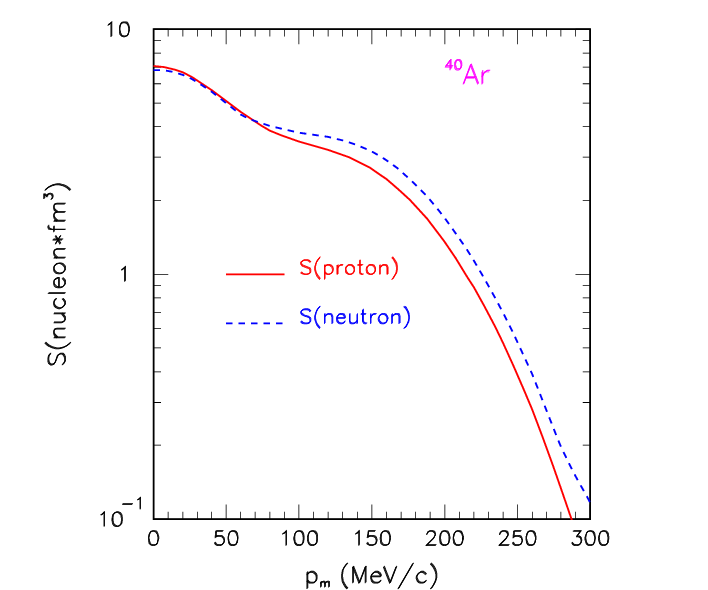}
  \end{center}
  \caption{\label{Fig1} Total proton and neutron momentum distributions in
    ${}^{40}$Ar nucleus.
}
\end{figure*}
orbitals of ${}^{40}$Ca and ${}^{40}$Ar are approximately 87\% on average.
Proton and neutron binding energies and the occupancy's of the orbitals in
${}^{40}$Ar are given in Table II of Ref.~\cite{BAV4}. In this work we assume
that the missing strength can be attributed to the short-range nucleon-nucleon
$(NN)$ correlations, leading to the appearance of the high-momentum and
high-energy nucleon distribution in the target. 
 
Figures 1 and 2 show the proton momentum distributions for occupied orbitals
in ${}^{12}$C and ${}^{16}$O, calculated within the mean-field approach. The
neutron momentum distributions in these nuclei are almost identical to proton
ones. The total proton and neutron momentum distributions in ${}^{40}$Ar are
presented in Fig.3. These distributions are normalized to the total number of
protons/neutrons on the IPSM shells.

For an outgoing nucleon, the simplest chose is to use plane-wave function $\Psi$
in Eq.(\ref{Eq.11}) that is, no interactions are between the ejected nucleon
$N$ and the residual nucleus $B$, i.e. to use the so-called plane-wave impulse
approximation (PWIA). For a more realistic description, final state interaction
effects should be taken into account. In the RDWIA the distorted-wave function
of the knocked out nucleon $\Psi$ is evaluated as a solution of a Dirac
equations containing a phenomenological relativistic optical potential~
\cite{Fissum}. This potential consists of a real part, which describes the
rescattering of the ejected nucleon and an imaginary part for the absorption
of it into unobserved channels. We use the LEA program~\cite{LEA} for numerical
calculation of the distorted-wave function with the EDAD1 parameterization~
\cite{Cooper} of the relativistic optical potential for carbon, oxygen and
calcium.

\section{Results and analysis}

The reduced cross sections of ${}^{12}$C$(e,e^{\prime}p)$ reaction in the range
of missing energy, that corresponds to knockout of $1s$ and $1p$-shells protons
were measured at Tokyo~\cite{Tokyo}, Saclay~\cite{SaclayC1, SaclayC2}, NIKHEF~
\cite{NIKHEFC}, SLAC~\cite{SLAC}, and JLab~\cite{Dutta}. The knockout of
$1p$-shell protons in ${}^{16}$O$(e,e^{\prime}p)$ was studied at Saclay~
\cite{SaclayC1, SaclayO}, NIKHEF~\cite{NIKHEFO1, NIKHEFO2}, Mainz~\cite{Mainz},
and JLab~\cite{Fissum}. In these experiments, cross sections data for the
lowest-lying fragments of each shell were measured as functions of $p_m$, and
normalization factors (relating how much the measured cross section data were
less than predicted in IPSM) were extracted. The E12-14-012
experiment~\cite{JLabE},
performed in JLab has measured the $(e,e^{\prime}p)$ reduced cross sections
using ${}^{40}$Ar~\cite{JLabAr} and  ${}^{48}$Ti~\cite{JLabTi} targets. The
reduced cross sections measured in the missing momentum and missing energy
ranges $15\leq p_m \leq 300$ MeV/c and  $12 \leq E_m \leq 80$ MeV.

The distorted spectral function depends upon initial momentum $\p_m$, ejectile
momentum $\p_x$ and angle between the initial and final nucleon momenta. Thus
it depends upon
kinematical conditions and is different for parallel and perpendicular
kinematics.
\begin{figure*}
  \begin{center}
    \includegraphics[height=10cm, width=18cm]{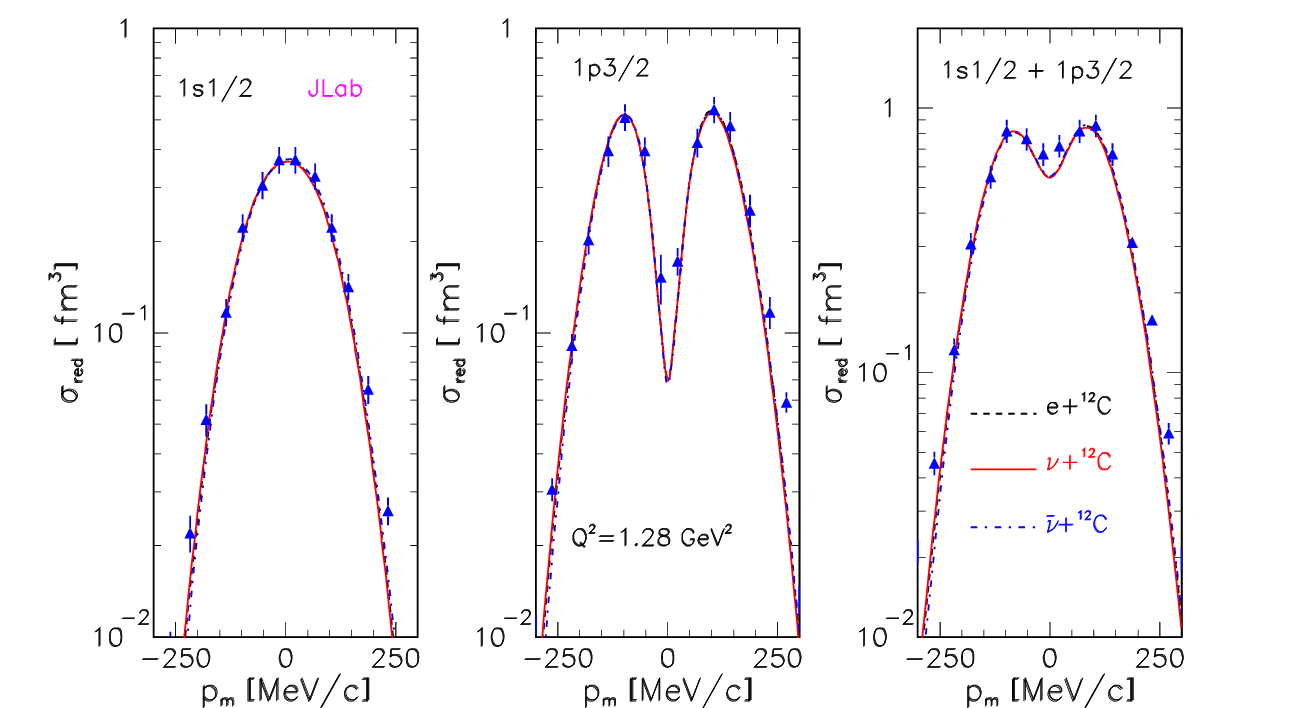}
  \end{center}
  \caption{\label{Fig4}
Comparison of the RDWIA calculations 
for electron, neutrino and antineutrino reduced  cross sections for the 
removal of nucleons from 1$s$ and 1$p$ shells of $^{12}$C as functions of the 
missing momentum. JLab data~\cite{Dutta} for beam energy $E_{beam}$=2.455 GeV, 
proton kinetic energy $T_p$=350 MeV, and $Q^2$=0.64 (GeV/$c)^2$. The RDWIA
calculations are shown for electron scattering (dashed line) and neutrino
(solid line) and antineutrino (dashed-dotted line) scattering. This figure
taken from Ref.~\cite{BAV2}.
}
\end{figure*}
Furthermore, $\sigma_{red}$ depends upon initial electron energy due
to Coulomb distortion. The RDWIA approach with LEA code was successfully
tested against measured ${}^{12}$C$(e,e^{\prime}p)$~\cite{Kelly1},
${}^{16}$O$(e,e^{\prime}p)$~\cite{Fissum}, and ${}^{40}$Ca$(e,e^{\prime}p)$~
\cite{BAV4} differential and reduced cross sections, and the normalization
factors $S(\alpha)$ for the IPSM orbitals were derived.

In Refs.~\cite{BAV1, BAV2, BAV4} electron and CCQE (anti)neutrino scattering on
oxygen, carbon, calcium, and argon targets were studied. It was found that the
reduced cross sections for (anti)neutrino scattering are similar to those of
electron scattering, and the latter are in good agreement with electron data.
The difference between the electron and (anti)neutrino reduced cross sections
calculated for Saclay kinematics is less than 10\%.
\begin{figure*}
  \begin{center}
    \includegraphics[height=10cm, width=18cm]{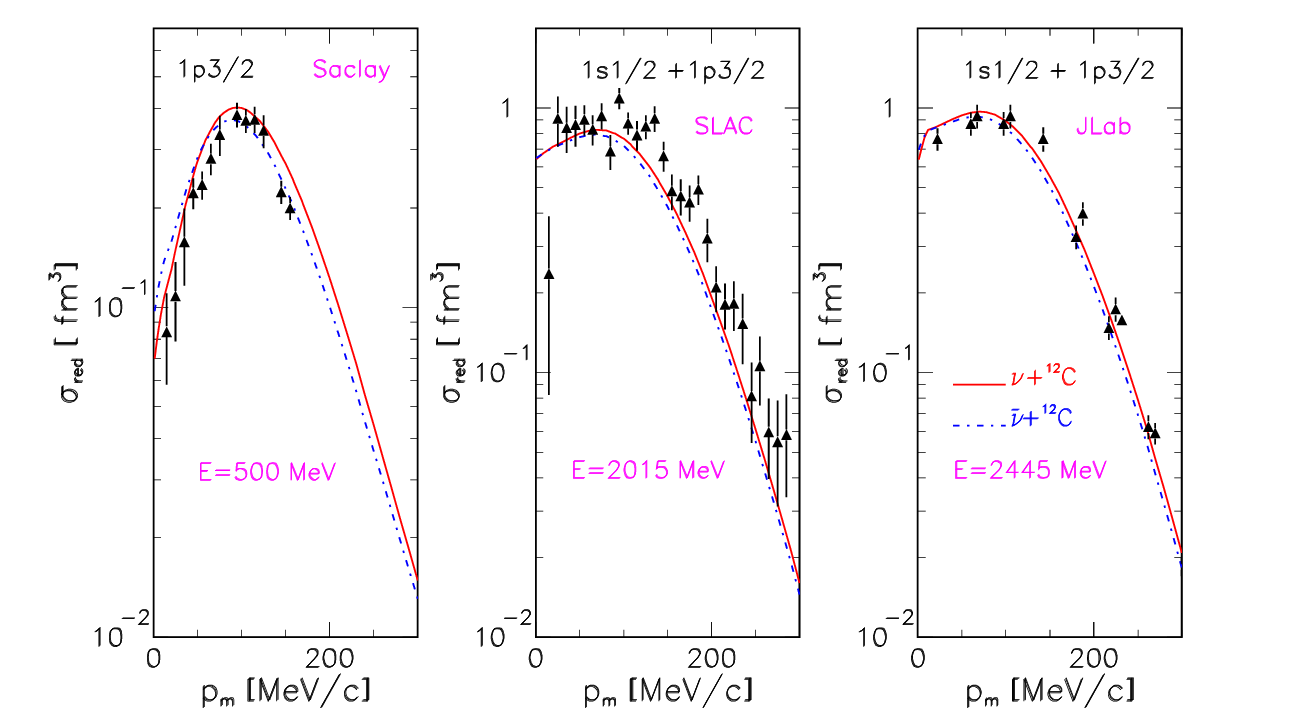}
  \end{center}
  \caption{\label{Fig5}
The RDWIA calculation of neutrino (solid line) and antineutrino 
(dashed-dotted line) averaged reduced cross 
sections compared with measured exclusive cross section data for
the removal of nucleons from 1$p$ and 1$s+1p$ shells of $^{12}$C as functions
of the missing momentum. The data are from Saclay~\cite{SaclayC1} for $1p$
and the beam energy $E_{beam}=500$ MeV, SLAC~\cite{SLAC} for $1s+1p$ shells
and $E_{beam}=2015$ MeV, and from JLab~\cite{Dutta} for $1s+1p$ shells and
$E_{beam}$=2455 MeV
}
\end{figure*}
This can be attributed to
Coulomb distortion upon electron wave function which is usually
described as the
effective momentum approximation (EMA)~\cite{EMA}.
In the EMA, the electron Coulomb wave function is replaced by a plane wave
function with effective momentum whose value is larger than the value of
electron momentum at infinity, because of Coulomb attraction. The flux is also
increased in the interaction zone by focusing of electron wave. This effect is
proportional to charge of the target and weakens as the beam energy increases.
The small difference between neutrino and antineutrino reduced cross sections
is due to the difference in the FSI of the proton and neutron with the residual
nucleus.

In this section we present the results of the RDWIA calculations of
the averaged reduced cross sections Eq.\eqref{avsec} for (anti)neutrino
scattering off carbon, oxygen, and argon as functions of the missing momentum
$p_m$ and compare them with the measured $(e,e^{\prime}p)$ reduced cross
sections.
 In Ref.~\cite{BAV2} electron, neutrino, and antineutrino cross sections for the
removal protons from the $1s$, $1p$, and $1s+1p$ shells of ${}^{120}$C as
functions of missing momentum $p_m$ were calculated and compared with JLab
data~\cite{Dutta}. For illustration, Fig.~\ref{Fig4} shows measured the removal
 cross sections as compared with the LEA code calculations~\cite{BAV2}.
\begin{figure*}
  \begin{center}
    \includegraphics[height=9cm,width=11cm]{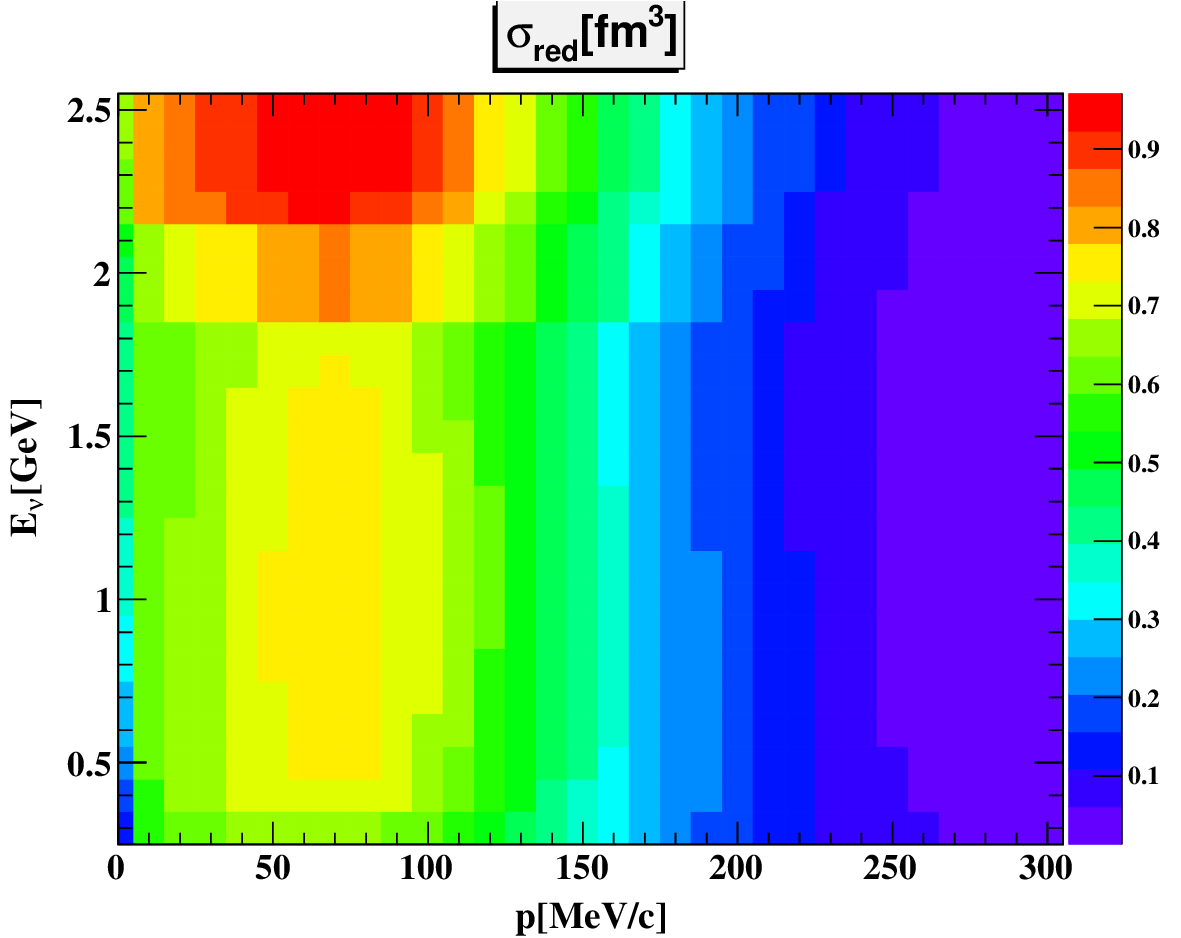}
  \end{center}
  \caption{\label{Fig6} The RDWIA neutrino averaged reduced cross section for
    removal of nucleons from the $1s+1p$ shells of ${}^{12}$C as a function of
    neutrino energy and missing momentum $p_m$.
}
\end{figure*}
It should be note that negative value
of $p_m$ corresponds to $\phi=\pi$ and positive to $\phi=0$, where $\phi$ is
the angle between the scattering $(\k_i,\k_f)$ and reaction $(\p_x,\p_B)$
planes. The data  for beam energy $E_{beam}=2.445$ GeV and $Q^2=0.64$ (GeV/c)$^2$
were measured in the quasi-perpendicular kinematics with constant $(\omega,\q)$.
The electron and neutrino scattering off the nuclei are closely interrelated
and one can treat both processes within the same formalism. There is an overall
 good agreement between the cross sections calculated in the RDWIA and data.

The averaged reduced cross sections for removal of nucleons from the $1p$, and
$1s+1p$ shells of ${}^{12}$C$(\nu_{\mu},\mu p)$ and
${}^{12}$C$(\bar{\nu}_{\mu},\mu n)$
reactions are shown in Fig.~\ref{Fig5} as functions of positive $p_m$  values
together with Saclay~\cite{SaclayC2}, SLAC~\cite{Kelly1}, and JLab~\cite{Dutta}
data.
\begin{figure*}
  \begin{center}
    \includegraphics[height=16cm,width=16cm]{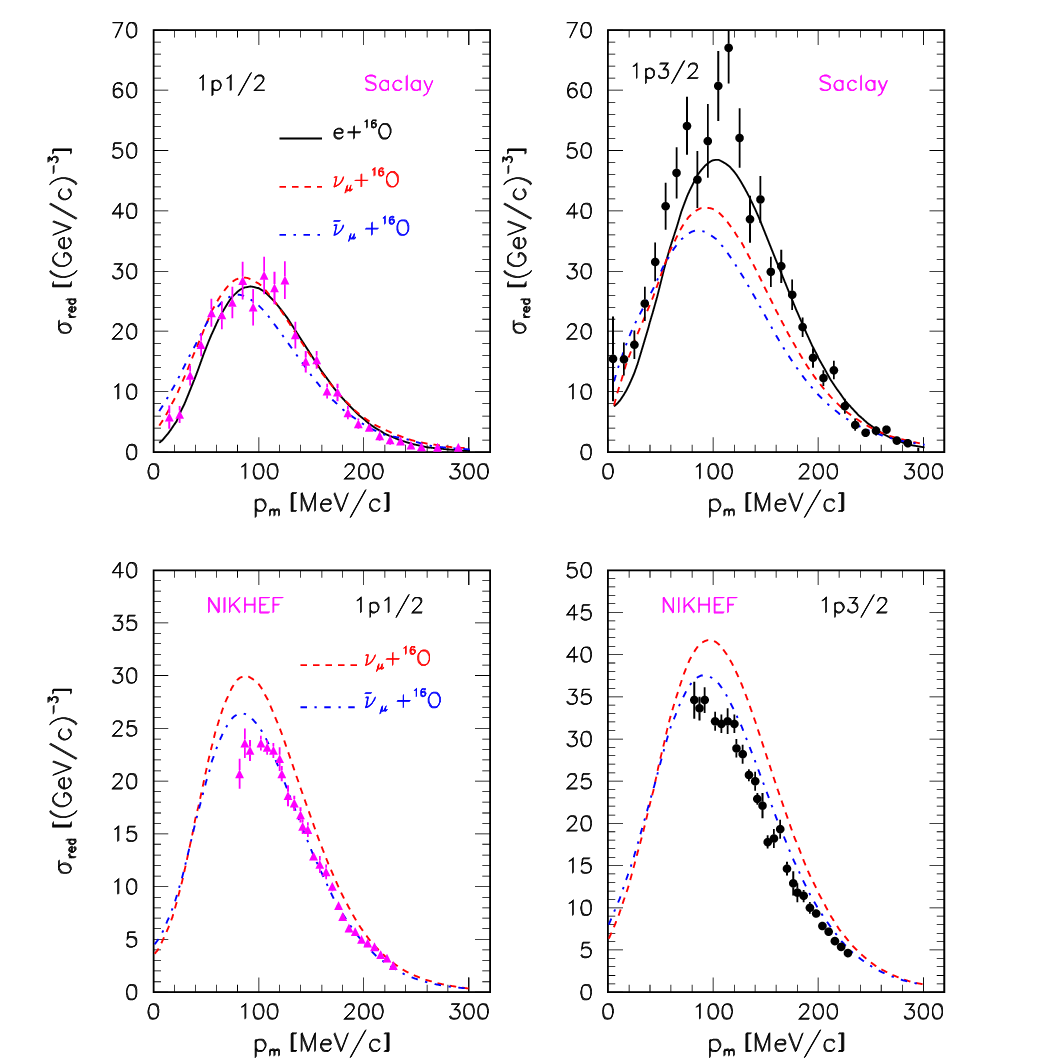}
  \end{center}
  \caption{\label{Fig7}
    Comparison of the RDWIA calculations for neutrino (dashed line) and
    antineutrino (dashed-dotted line) averaged reduced  cross sections for the 
removal of nucleons from 1$p$ shell of $^{16}$O with Saclay~\cite{SaclayC1} and
NIKHEF~\cite{NIKHEFO1} data as functions of $p_m$. Also shown are the RDWIA
calculations of the reduced cross section for electron scattering (solid line)
from Ref.~\cite{BAV2}.
}
\end{figure*}
The data for beam energies $E_{beam}=500, 2015$, and 2445 MeV were
measured. There is an overall agreement between the calculated averaged cross
sections and reduced cross sections of the $(e,e^{\prime}p)$ reaction measured
in different kinematics. The RDWIA averaged reduced cross section for removal
nucleon from $1s+1p$ shells in ${}^{12}$C$(\nu_{\mu},\mu p)$ is shown in
Fig.~\ref{Fig6} as a function of incoming neutrino energy and missing momentum
$p_m$. In the range of the maximum at $60 \leq p_m \leq 90$ MeV/c the cross
section increases slowly with neutrino energy, and slight changes at
$p_m \geq 120$ MeV/c and $p_m \leq 40$ MeV/c.    

The averaged reduced cross sections for the removal of nucleons from the
$1p$ shell
in ${}^{16}$O$(\nu_{\mu},\mu p)$ and ${}^{16}$O$(\bar{\nu}_{\mu}, \mu n)$
reactions are shown in Fig.~\ref{Fig7} as functions of $p_m$ together with
Saclay~\cite{SaclayC1} and NIKHEF~\cite{NIKHEFO1} data. There is an overall
agreement between calculated cross sections and data, but the values of
the calculated cross sections at maximum is systematically higher (about 15-20\%)
than measured ones for NIKHEF kinematics. Unfortunately, there are no data for
removal protons from the $1s$ and $1s+1p$ shells of ${}^{16}$O. Therefore the
models of lepton-nucleus interaction that don't take into account the shell
structure of nucleus can not be tested against the available reduced cross
sections measured in ${}^{16}$O$(e,e^{\prime}p)$ reaction.
\begin{figure*}
  \begin{center}
    \includegraphics[height=9cm,width=11cm]{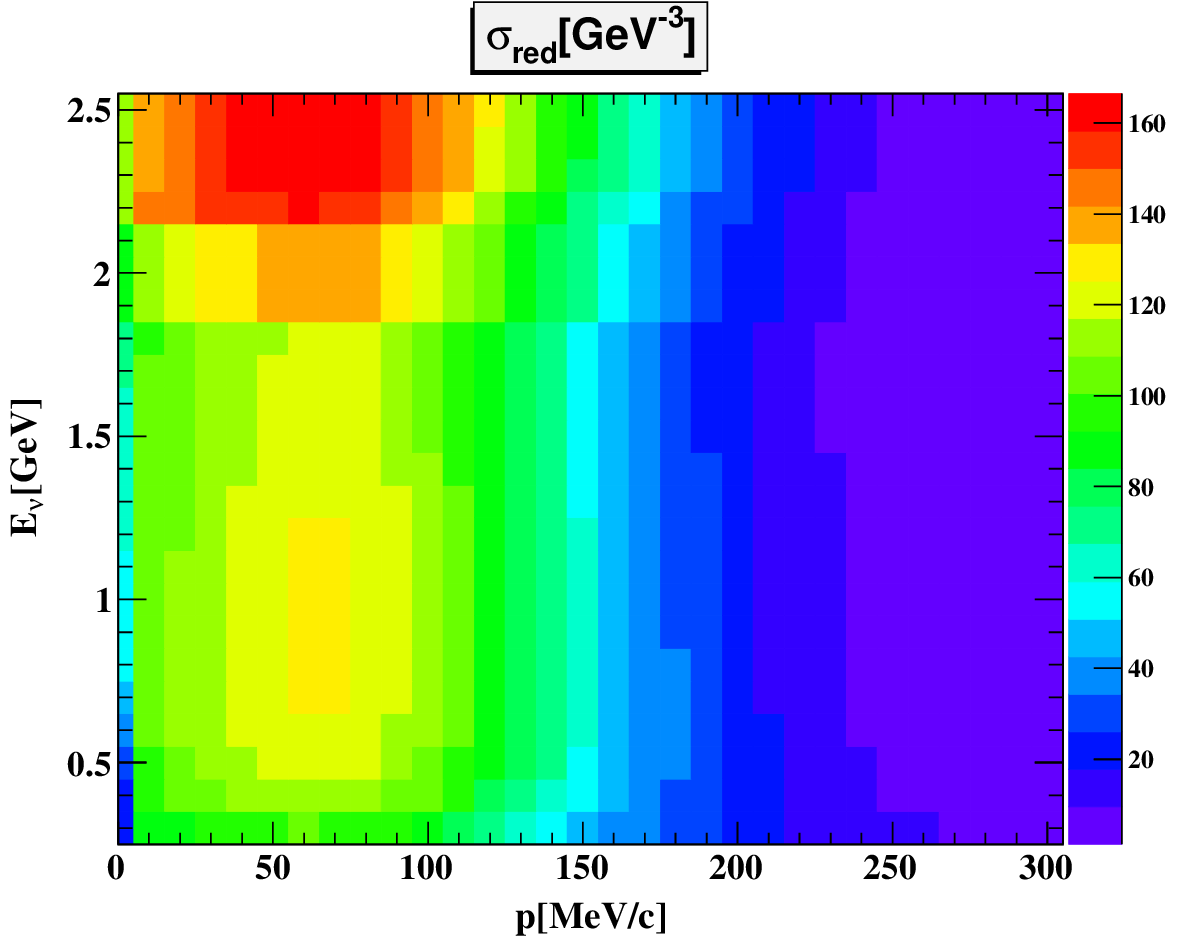}
  \end{center}
  \caption{\label{Fig8} Same as Fig.~\ref{Fig6} but for the ${}^{16}$O.
}
\end{figure*}
The RDWIA averaged reduced cross section for removal nucleon from $1s+1p$
shells in ${}^{16}$O$(\nu_{\mu},\mu p)$ is shown in Fig.~\ref{Fig8} as a function
of incoming neutrino energy and missing momentum $p_m$. As can be seen from
figures~\ref{Fig6} and~\ref{Fig8} the dependence of these cross sections upon
neutrino energy  and $p_m$ are almost similar.
\begin{figure*}
  \begin{center}
    \includegraphics[height=16cm,width=16cm]{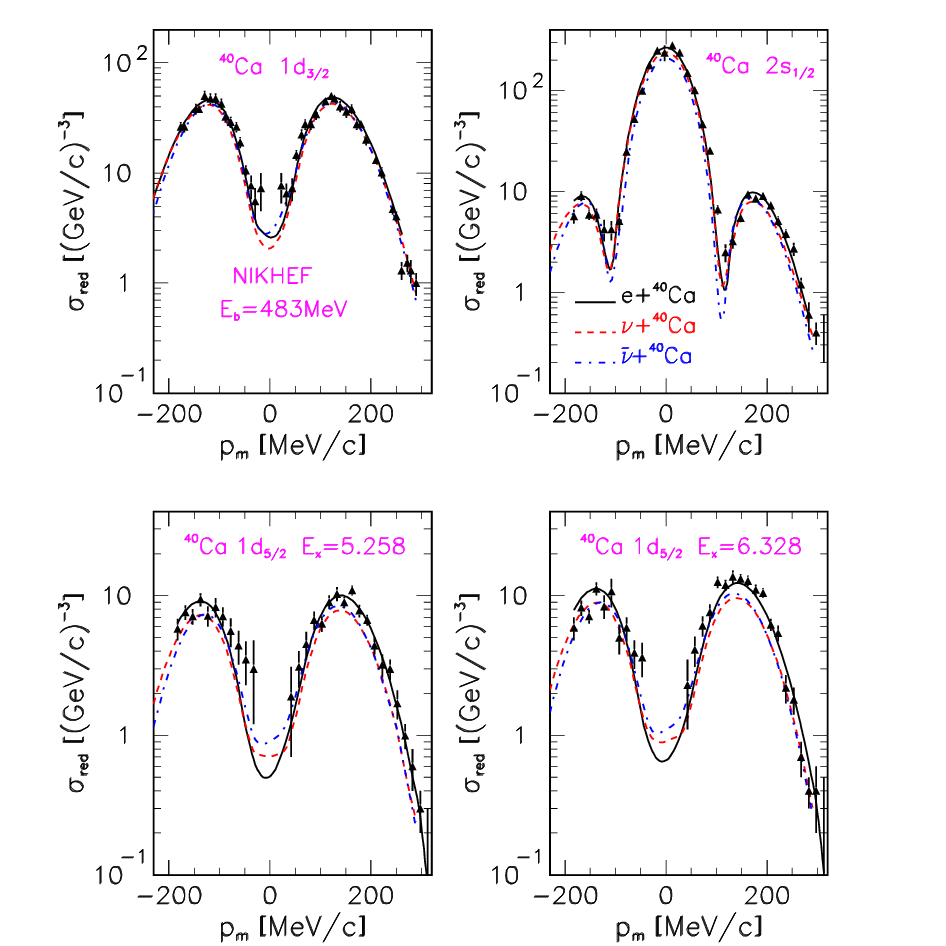}
  \end{center}
  \caption{\label{Fig9} Comparison of the RDWIA calculations for electron 
(solid line), neutrino (dashed line), and antineutrino (dashed-dotted line) 
    reduced  cross sections for the removal of nucleons from  1$d_{3/2}$,
    2$s_{1/2}$, and 1$d_{5/2}$ shells of $^{40}$Ca with NIKHEF data~
    \cite{Kramer3}. The cross sections are presented as functions of missing
    momentum $p_m$. The figure taken from~Ref.~\cite{BAV4}.
}
\end{figure*}

The structure of calcium and argon nuclei is similar, although unlike
${}^{40}_{18}$Ar, ${}^{40}_{20}$Ca is a symmetric and closed-shell nucleus.
\begin{figure*}
  \begin{center}
    \includegraphics[height=8cm,width=19cm]{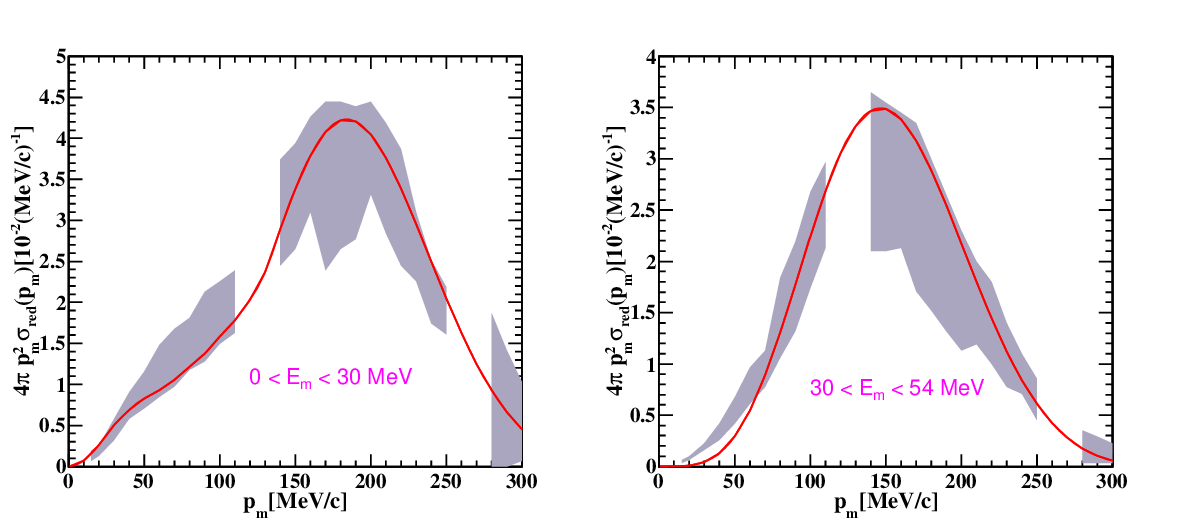}
  \end{center}
  \caption{\label{Fig10} Missing momentum distribution in argon obtained by
    integrating over the missing energy range of 0 - 30 MeV (left panel) and
    30 - 54 MeV (right panel), presented with the geometrical factor of
    $4\pi p_m^2$. The gray band shows the measured spectral function including
     the full error.
}
\end{figure*}
In Ref.~\cite{BAV4} the $(e,e^{\prime}p)$ reduced cross sections for removal of
the proton from $1d_{3/2}$ shell, for transition to the 1/2$^{+}$ exited state of
the ${}^{39}$K nucleus at excitation energy $E_x=2.522$ MeV, and for the
transitions to the 5/2$^{+}$  excited states at $E_x$=5.258 MeV and
$E_x$=6.328 MeV, obtained by knocking out protons from the 2$s_{1/2}$ and
1$d_{5/2}$ orbitals, correspondingly were calculated. The calculated reduced
cross sections are shown in Fig.~\ref{Fig9} with the NIKHEF
data~\cite{Kramer1, Kramer2} and provide a good description of the shape and
magnitude of the measured distribution. Neutrino and antineutrino calculated
reduced cross sections of ${}^{40}$Ca$(\nu,\mu^{-}p){}^{39}$Ca and 
${}^{40}$Ca$(\bar{\nu},\mu^{+}n){}^{39}$K reactions also shown are in
Fig.~\ref{Fig9}. There is an overall good agreement between calculated cross
sections, but the values of the electron cross sections at the maximum is
systematically higher than those for (anti)neutrino. This can be attributed to
Coulomb distortion upon the incident electron wave function.

The JLab experiment~\cite{JLabE} has measured the $(e,e^{\prime}p)$ cross
sections using argon and titanium targets ~\cite{JLabAr, JLabTi}. The reduced
cross sections were obtained in the missing momentum $15 \leq p_m \leq 300$
MeV/c and missing energy range $12 \leq p_m \leq 80$ MeV.
\begin{figure*}
  \begin{center}
    \includegraphics[height=8cm,width=12cm]{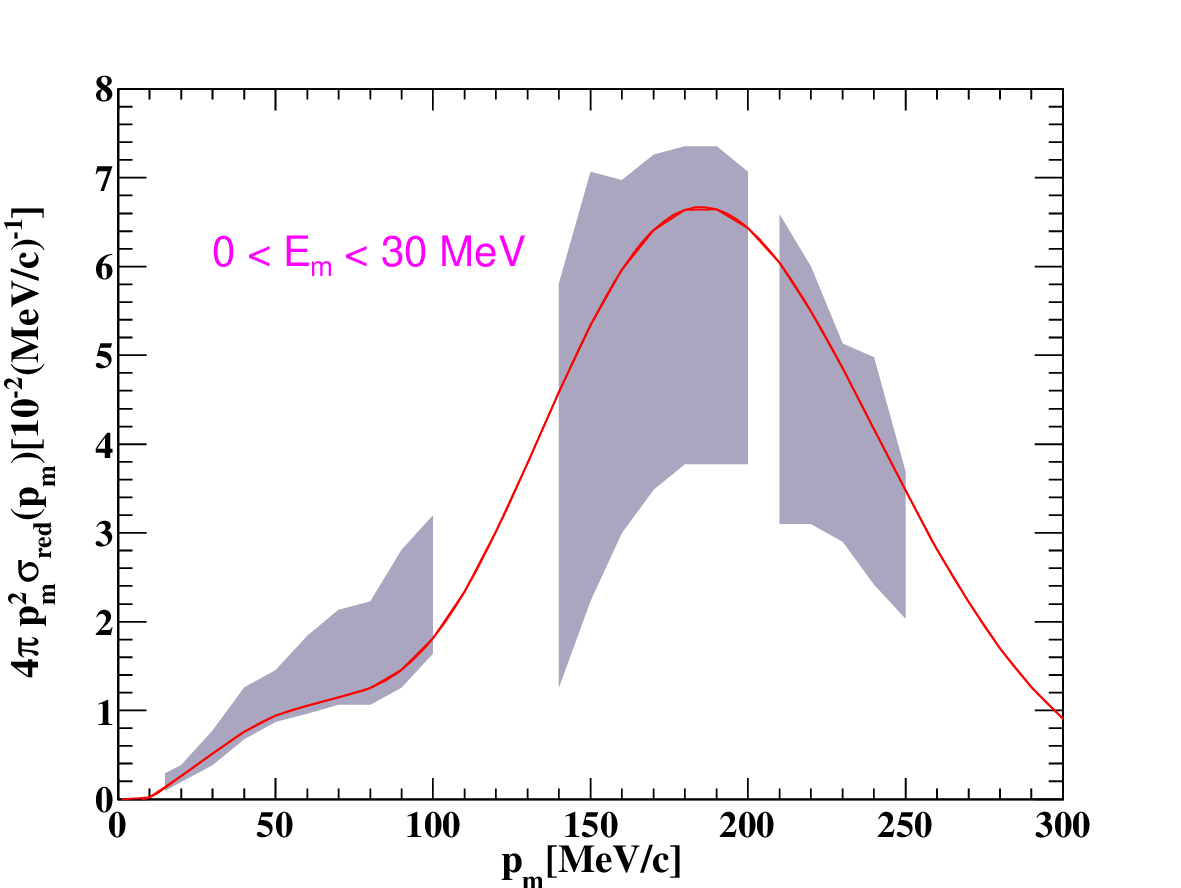}
  \end{center}
  \caption{\label{Fig11} Missing momentum distribution in titanium obtained by
    integrating over the missing energy range of 0 - 30 MeV, presented with
    the geometrical factor of $4\pi p_m^2$. The gray band shows the measured
    spectral function including the full error.
}
\end{figure*}
The procedure to obtained information on neutron
distribution in argon is based on the observation that neutron spectrum of
${}^{40}_{18}$Ar is mirrored by proton spectrum of the nucleus of titanium,
having charge $Z=22$. Therefor one can expect that the proton spectral function
obtained from Ti$(e,e^{\prime}p)$ data provides information on the neutron
spectral function of argon.

The ${}^{40}_{18}$Ar and ${}^{48}_{22}$Ti data were analyzed to obtain the
spectral functions, describing the energy and momentum distributions of protons
in the argon and titanium ground states. The effect of FSI, which are known to
be significant in $(e,e^{\prime}p)$ reactions, was taken into account within the
distorted-wave impulse approximation approach. Figure~\ref{Fig10} shows the
missing momentum distributions of protons in argon obtained by integrating the
data over the missing energy ranges $0 - 30$ MeV and $30 - 50$ MeV. The proton
missing momentum distribution in titanium was obtained by integrating the data over the
missing energy range $0 - 30$ MeV is shown in Fig.~\ref{Fig11}. Also shown in
Figs.~\ref{Fig10} and~\ref{Fig11} are the results obtained without FSI effects
in the relativistic plane wave impulse approximation (RPWIA), with
normalization factors $S_{\alpha}$ from Ref.~\cite{BAV4}. 
There is an overall agreement between the RPWIA calculations and data within
sizable uncertainties of the measured proton momentum distributions. A more
accurate determination of the distorted spectral functions for different shells
of ${}^{40}$Ar and ${}^{48}$Ti will improve the testing of models using for
description of neutrino interaction with these nucleus.

\begin{figure*}
  \begin{center}
    \includegraphics[height=9cm,width=11cm]{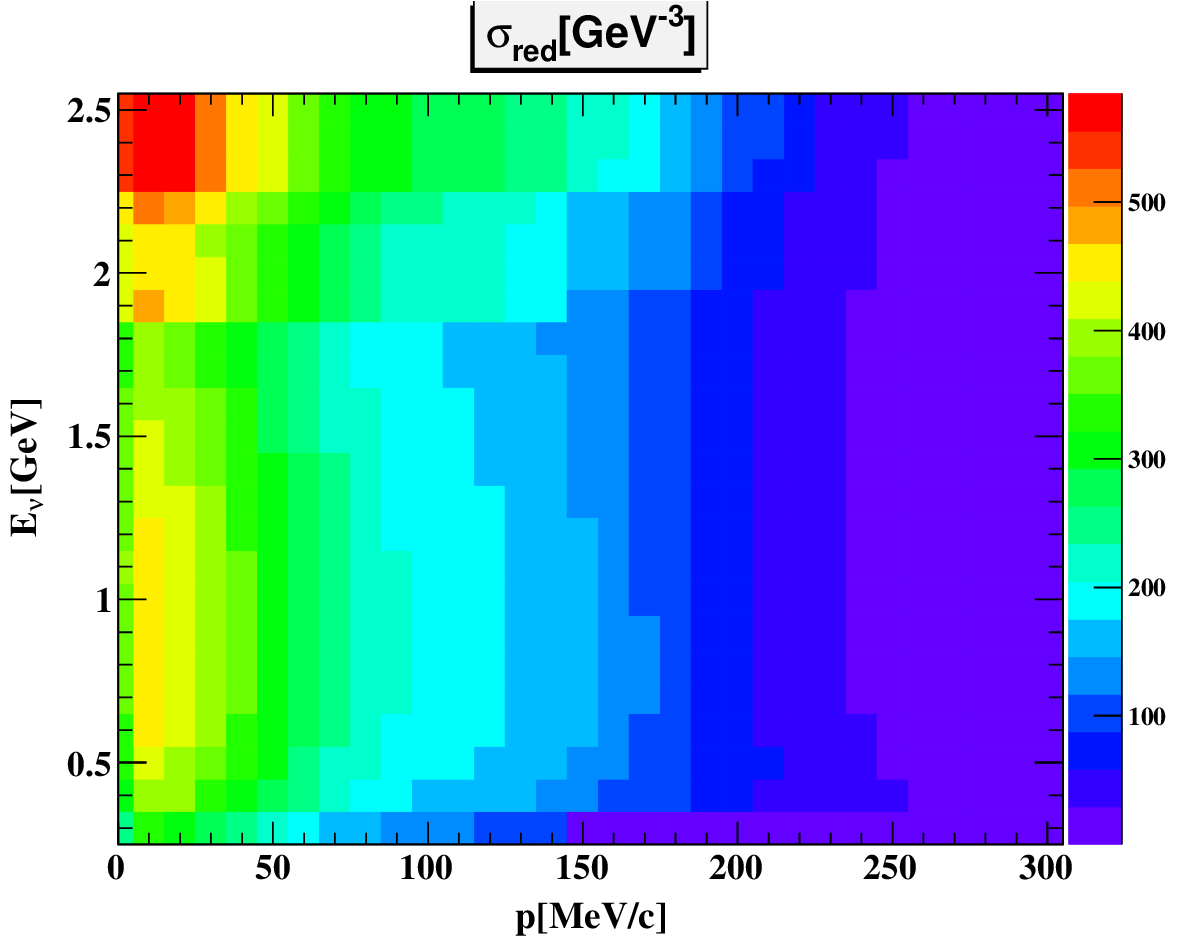}
  \end{center}
  \caption{\label{Fig12} Same as Fig.~\ref{Fig6} but for the ${}^{40}$Ar.
}
\end{figure*}
The averaged reduced cross section of ${}^{40}$Ar$(\nu_{\mu},\mu p)$
reaction calculated in the RDWIA approach is shown in Fig.~\ref{Fig12} as
a function of neutrino energy and missing momentum $p_m$. It has to be pointed
out that unlike ${}^{12}$C and ${}^{16}$O, in ${}^{40}$Ar the maximum of these
cross sections is shifted to the range of lower missing momentum
$p_m\approx 15$ MeV/c. The cross section increases very slowly with neutrino
energy. 

Neutrino event generators employ the factorization approach to make predictions
about the lepton and also the outgoing nucleon kinematics from inclusive
models. These models are aimed to describe inclusive cross section 
that is only as function of the final lepton kinematics. This factorization 
uses the spectral functions, which are generated from different nucleon
distributions in the initial nuclear state (local Fermi 
gas, shell model, etc). While the behavior of the cross section against the 
lepton kinematics may be described correctly, there is no guarantee that the 
correlations between the final lepton and nucleon for a given event are 
preserved. The comparison of the employed spectral function with the measured 
reduced cross sections allows the estimation of the accuracy of the nuclear effects 
calculations. On the other hand, the effective spectral functions can be 
obtained within the microscopic and unfactorized models, like the RDWIA, that 
successfully describes exclusive $(e,e^{\prime}p)$ cross sections and employed in
 the neutrino event generators.

\section{Conclusions}

In this article, we studied within the RDWIA approach the semi-exclusive
reduced cross sections of CCQE (anti)neutrino scattering on carbon, oxygen,
and argon. We calculated averaged over phase space reduced cross sections for
removal of nucleons from the $\mathrm{1p}$,  and $\mathrm{1s+1p}$ shells of
${}^{12}$C$(\nu_{\mu},\mu p)$, ${}^{12}$C$(\bar{\nu}_{\mu},\mu n)$
and ${}^{16}$O$(\nu_{\mu},\mu p)$, ${}^{16}$O$(\bar{\nu}_{\mu},\mu n)$ reactions
as functions of missing momentum and incoming neutrino energy, and compared
them with the reduced cross sections obtained from measurements of
$(e,e^{\prime}p)$ scattering on ${}^{12}$C and ${}^{16}$O. We also calculated in
the relativistic plane wave impulse approximation the averaged reduced cross
sections for single nucleon knocked out in ${}^{40}$Ar$(\nu_{\mu},\mu p)$ and
${}^{49}$Ti$(\bar{\nu}_{\mu},\mu n)$ reactions as functions of $p_m$ and
$\varepsilon_{\nu}$ and compared them with proton momentum distributions in
argon and titanium obtained from $(e,e^{\prime}p)$ scattering in the JLab
experiments

We found that the shape and magnitude of the averaged reduced cross sections
for (anti)neutrino scattering as a function of missing momentum are similar to
measured reduced cross sections of electron scattering. The averaged removal
cross sections calculated for argon and titanium within RPWIA approach seem
mostly consistent with the data within sizable uncertainties of measured proton
momentum distributions. The difference less than 10\% between the electron and
(anti)neutrino cross sections can be attributed to Coulomb distortion upon
incoming electron wave function. The small difference between neutrino and
antineutrino reduced cross sections is due to difference in the FSI of proton
and neutron with the residual nucleus. The averaged reduced cross sections for
removal nucleon from $\mathrm{1s+1p}$ shells in carbon and oxygen have maximum
in the range of missing momentum $60 \leq p_m \leq 90$ MeV/c and in ${}^{40}$Ar
the maximum is shifted in the range of $p_m\approx 15$ MeV/c. The cross
sections increase very slowly with neutrino energy.

Some neutrino event generators employ the factorization approach to make
prediction about the lepton and outgoing nucleon kinematics, using different
nucleon distributions in the ground nuclear state. In this way the direct
comparison of the implemented spectral functions with the precise electron
reduced cross sections data allows to estimate the accuracy of the nuclear
effects calculations, such as the nuclear ground state and FSI.

\section*{Acknowledgments}

The author greatly acknowledge A. Habig for fruitful discussions and a
critical reading of the manuscript. A would like to thank S. Luchuk for his
constructive comments and suggestions.
 
%


\end{document}